%% file: main.tex
\pdfoutput=1
\documentclass[conference]{IEEEtran}

\usepackage[utf8]{inputenc}
\usepackage[english]{babel}

\usepackage{csquotes}

\usepackage{amsmath,amssymb,wasysym}

\usepackage{graphicx,color}
\usepackage[dvipsnames]{xcolor}
\usepackage{float}

\usepackage{tikz}
\usetikzlibrary{automata,arrows,calc,matrix,shadows}

\usepackage{tabularx}
\usepackage{makecell}
\usepackage{multirow}
\usepackage{colortbl}

\usepackage{listings}
\lstdefinestyle{vhdl_style}
{
    language=VHDL,
    float=!htb,
    basicstyle=\ttfamily\footnotesize,
    identifierstyle=\bfseries\color{black},
    keywordstyle=\bfseries\color{OrangeRed!80},
    stringstyle=\bfseries\color{yellow},
    commentstyle=\bfseries\color{gray},
    columns=flexible,
    frame=single,
    showspaces=false,
    showstringspaces=false,
    numberstyle=\tiny,
    stepnumber=1,
    breaklines=true,
    xrightmargin=-\fboxsep,
    backgroundcolor=\color{white},
    captionpos=t,
    mathescape,
    escapechar=\%
}

\usepackage{algorithm}
\usepackage[noend]{algpseudocode}
\algrenewcommand\algorithmicrequire{\textbf{\ \ Input:}}
\algrenewcommand\algorithmicensure{\textbf{Output:}}

\usepackage{url}
\usepackage{hyperref}

\usepackage{environ}

\usepackage{microtype}
\usepackage{xspace}

\usepackage{comment}

\usepackage[nolist]{acronym}
\input{acronyms.tex}

\newcommand{\circled}[2][]{
  \tikz[baseline=(char.bas<<e)]{
    \node[shape=circle,draw,inner sep=1pt,fill=black]
    (char) {\phantom{\ifblank{#1}{#2}{#1}}};
    \node[text=white] at (char.center) {\makebox[0pt][c]{\bf #2}};}}
\robustify{\circled}

\makeatletter
    \let\matamp=&
    \catcode`\&=13
\makeatletter
    \def&{
        \iftikz@is@matrix
            \pgfmatrixnextcell
        \else
            \matamp
        \fi
    }
\makeatother

\newcounter{lines}

\NewEnviron{timeline}[0]{
    \setcounter{lines}{1}
    \begin{tikzpicture}[
        column 1/.style={anchor=east},
        column 2/.style={anchor=west, text width=0.85\linewidth},
        text depth=0pt,
        text height=1ex,
        row sep=2ex,
        column sep=1em,
        ]

        \matrix(m)[matrix of nodes]{\BODY};
        \pgfmathtruncatemacro\endmtx{\thelines-1}

        \path[fill=black!75, draw=black!75]
        ($(m-1-1.north east)!0.5!(m-1-2.north west)$)--($(m-\endmtx-1.south east)!0.5!(m-\endmtx-2.south west)$);

        \foreach \x in {1,...,\endmtx} {
            \node[circle, fill=black!75, draw=black!75, inner sep=0.15pt, draw=white, thick]
            (m-c-\x) at
            ($(m-\x-1.east)!0.5!(m-\x-2.west)$){};
            \draw[fill=black!75, draw=black!75] (m-c-\x.west)--++(-3pt,0);
         }

        \draw[fill=black!75, draw=black!75]([yshift=2pt]m.north west)--([yshift=2pt]m.north east);
        \draw[fill=black!75, draw=black!75]([yshift=-2pt]m.south west)--([yshift=-2pt]m.south east);
    \end{tikzpicture}
}

\newcommand{\etal}[0]{\textit{et al.}\xspace}

\newcommand{\romannumber}[1]{\uppercase\expandafter{\romannumeral#1}}

\newcommand{\Section}[1]{\textcolor{blue}{Section~\ref{#1}}}

\begin{document}

\title{
    Hardware Reverse Engineering:\\Overview and Open Challenges
}

\author{
    \IEEEauthorblockN{
        Marc Fyrbiak\IEEEauthorrefmark{1},
        Sebastian Strauß\IEEEauthorrefmark{2},
        Christian Kison\IEEEauthorrefmark{1},
        Sebastian Wallat\IEEEauthorrefmark{3},
        Malte Elson\IEEEauthorrefmark{2},\\
        Nikol Rummel\IEEEauthorrefmark{2},
        Christof Paar\IEEEauthorrefmark{1}\IEEEauthorrefmark{3}, \IEEEmembership{Fellow,~IEEE}\\
    }
    \IEEEauthorblockA{
        \IEEEauthorrefmark{1}Horst G\"ortz Institute for IT Security, Ruhr University Bochum, Germany\\
        \IEEEauthorrefmark{2}Institute of Educational
        Research, Ruhr University Bochum, Germany\\
        \IEEEauthorrefmark{3}University of Massachusetts Amherst, USA\\
        \{\textit{prename.surname}\}@rub.de,
        \{\textit{swallat}\}@umass.edu\\
    }
}

\maketitle

\input{section/abstract.tex}


\input{section/introduction}
\input{section/hardware_reverse_engineering}
\input{section/challenges_human_factor}
\input{section/conclusion}

\section*{Acknowledgments}
This work is partially supported by ERC grant No. 695022.

\bibliographystyle{IEEEtran}
{\small
        \bibliography{bibliography}
}

\end{document}

%% file: acronyms.tex
\begin{acronym}

\acro{3DES}{Triple-DES}

\acro{AES}{Advanced Encryption Standard}
\acro{ALU}{Arithmetic Logic Unit}
\acro{API}{Application Programming Interface}
\acro{ARX}{Addition Rotation XOR}
\acro{ASIC}{Application Specific Integrated Circuit}
\acro{ASIP}{Application Specific Instruction-Set Processor}
\acro{AS}{Active Serial}

\acro{BDD}{Binary Decision Diagram}
\acro{BGL}{Boost Graph Library}
\acro{BNF}{Backus-Naur Form}
\acro{BRAM}{Block-Ram}

\acro{CBC}{Cipher Block Chaining}
\acro{CFB}{Cipher Feedback Mode}
\acro{CFG}{Control Flow Graph}
\acro{CLB}{Configurable Logic Block}
\acro{CLI}{Command Line Interface}
\acro{COFF}{Common Object File Format}
\acro{CPA}{Correlation Power Analysis}
\acro{CPU}{Central Processing Unit}
\acro{CRC}{Cyclic Redundancy Check}
\acro{CTR}{Counter}

\acro{DC}{Direct Current}
\acro{DES}{Data Encryption Standard}
\acro{DFA}{Differential Frequency Analysis}
\acro{DFT}{Discrete Fourier Transform}
\acro{DLL}{Dynamic Link Library}
\acro{DMA}{Direct Memory Access}
\acro{DNF}{Disjunctive Normal Form}
\acro{DPA}{Differential Power Analysis}
\acro{DSO}{Digital Storage Oscilloscope}
\acro{DSP}{Digital Signal Processing}
\acro{DUT}{Device Under Test}

\acro{ECB}{Electronic Code Book}
\acro{ECC}{Elliptic Curve Cryptography}
\acro{EEPROM}{Electrically Erasable Programmable Read-only Memory}
\acro{EMA}{Electromagnetic Emanation}
\acro{EM}{electro-magnetic}

\acro{FFT}{Fast Fourier Transformation}
\acro{FF}{Flip Flop}
\acro{FI}{Fault Injection}
\acro{FIB}{Focused Ion Beam}
\acro{FIR}{Finite Impulse Response}
\acro{FPGA}{Field Programmable Gate Array}
\acro{FSM}{Finite State Machine}

\acro{GUI}{Graphical User Interface}

\acro{HDL}{Hardware Description Language}
\acro{HD}{Hamming Distance}
\acro{HF}{High Frequency}
\acro{HSM}{Hardware Security Module}
\acro{HW}{Hamming Weight}

\acro{IC}{Integrated Circuit}
\acro{IO}[I/O]{Input/Output}
\acro{IOB}{Input Output Block}
\acro{IOT}[IoT]{Internet of Things}
\acro{IP}{Intellectual Property}
\acro{ISA}{Instruction Set Architecture}
\acro{IV}{Initialization Vector}

\acro{JTAG}{Joint Test Action Group}

\acro{KAT}{Known Answer Test}

\acro{LFSR}{Linear Feedback Shift Register}
\acro{LSB}{Least Significant Bit}
\acro{LUT}{Look-up table}

\acro{MAC}{Message Authentication Code}
\acro{MIPS}{Microprocessor without Interlocked Pipeline Stages}
\acro{MMIO}{Memory Mapped \acl{IO}}
\acro{MSB}{Most Significant Bit}

\acro{NASA}{National Aeronautics and Space Administration}
\acro{NSA}{National Security Agency}
\acro{NVM}{Non-Volatile Memory}

\acro{OFB}{Output Feedback Mode}
\acro{OISC}{One Instruction Set Computer}
\acro{ORAM}{Oblivious Random Access Memory}
\acro{OS}{Operating System}

\acro{PAR}{Place-and-Route}
\acro{PCB}{Printed Circuit Board}
\acro{PC}{Personal Computer}
\acro{PS}{Passive Serial}
\acro{PUF}{Physical Unclonable Function}

\acro{RAM}{Random Access Memory}
\acro{RISC}{Reduced Instruction Set Computer}
\acro{RNG}{Random Number Generator}
\acro{ROI}{Region of Interest}
\acro{ROM}{Read-Only Memory}
\acro{ROP}{Return-oriented Programming}
\acro{RTL}{Register Transfer Level}

\acro{SCA}{Side-Channel Analysis}
\acro{SEM}{Scanning Electron Microscope}
\acro{SHA}{Secure Hash Algorithm}
\acro{SNR}{Signal-to-Noise Ratio}
\acro{SOD}{spin-on dielectric}
\acro{SPA}{Simple Power Analysis}
\acro{SPI}{Serial Peripheral Interface Bus}
\acro{SRAM}{Static Random Access Memory}

\acro{UART}{Universal Asynchronous Receiver Transmitter}
\acro{UHF}{Ultra-High Frequency}
\acro{USB}{Universal Serial Bus}

\acro{VHDL}{Very High Speed Integrated Circuit Hardware Description Language}

\acro{WISC}{Writeable Instruction Set Computer}

\acro{XDL}{Xilinx Description Language}
\acro{XTS}{XEX-based Tweaked-codebook with ciphertext Stealing}

\end{acronym}

%% file: section/abstract.tex

\begin{abstract}

Hardware reverse engineering is a universal tool for both legitimate and illegitimate purposes. On the one hand, it supports confirmation of \acs{IP} infringement and detection of circuit malicious manipulations, on the other hand it provides adversaries with crucial information to plagiarize designs, infringe on \acs{IP}, or implant hardware Trojans into a target circuit.
Although reverse engineering is commonplace in practice, the quantification of its complexity is an unsolved problem to date since both technical and human factors have to be accounted for. A sophisticated understanding of this complexity is crucial in order to provide a reasonable threat estimation and to develop sound countermeasures, i.e. obfuscation transformations of the target circuit, to mitigate risks for the modern \acs{IC} landscape.

The contribution of our work is threefold:
first, we systematically study the current research branches related to hardware reverse engineering ranging from decapsulation to gate-level netlist analysis. Based on our overview, we formulate several open research questions to scientifically quantify reverse engineering, including technical and human factors.
Second, we survey research on problem solving and on the acquisition of expertise and discuss its potential to quantify human factors in reverse engineering.
Third, we propose novel directions for future interdisciplinary  research encompassing both technical and psychological perspectives that hold the promise to holistically capture the complexity of hardware reverse engineering.

\end{abstract}

%% file: section/introduction.tex

\section{Introduction}
\label{hwre:section:introduction}

Reverse engineering refers to the process of information retrieval from a product, ranging from aircrafts to modern \acp{IC}, in order to understand its composition and inner workings~\cite{smc:1985:rekoff}. In a security context, it is often associated with analysis of proprietary binary programs~\cite{it:2012:willems,isc:2003:oorschot} or proprietary hardware chips~\cite{book:hardware_obfuscation:chapter1}. In particular, reverse engineering of the latter is a many-faceted process involving various methods and techniques such as decapsulation, delayering, imaging, and post-processing~\cite{quadir:2016:jetc}. Typically, several difficulties complicate the reverse engineering process in practice such as (1) lack of  meaningful descriptive information (e.g., names and comments), (2) lack of  module boundary information, and (3) lack of hierarchy of modules~\cite{subramanyan:2013:tect}.

Even though reverse engineering is a universal tool, in the hardware context it is often associated with illegitimate actions such as \ac{IP} infringement, weakening of security functions, or disclosure of necessary information for injecting hardware Trojans \cite{quadir:2016:jetc}. In fact, \ac{IP} infringement is a major concern for the industry. It is estimated that \ac{IC} companies face losses of several billion dollars in annual global revenue~\cite{ieee:2014:guin} due to reverse engineering.
In addition to commercial players, reverse engineering is also a major concern for governmental and military systems. Low-quality counterfeited hardware poses a devastating safety consequence for mission-critical systems such as airplanes, and weakened security systems result in the disclosure of classified information with all of its consequences.
In contrast, there are also various reasons to utilize reverse engineering for legitimate applications such as failure analysis, detection of counterfeit products~\cite{ieee:2014:guin} or hardware Trojans~\cite{bhunia:2014:ieee}. Furthermore, reverse engineering is also legal in many countries for competitive product analyses, education, and research, as long as copyrights and patents are not violated.

Despite intensive research on hardware reverse engineering~\cite{quadir:2016:jetc,vijayakumar:2017:tifs} and companies that perform on-demand reverse engineering~\cite{spectrum:2000:kumagai,torrance:2009:ches}, reverse engineering is still an opaque and poorly understood process. The question is not whether analysts are able to reverse engineer a given design, since with sufficient resources reverse engineering will always succeed. Rather, the fundamental research question is:
\begin{center}
	\enquote{\it How time-consuming and, thus, costly is the reverse engineering process of a proprietary design for successfully extracting crucial information?}
\end{center}
A sound quantification lays the foundation for reasonable threat estimates and development of sound countermeasures to mitigate the risks. Possible countermeasures include novel obfuscation strategies that hinder human analysts from reverse engineering based on a scientific evidence of both technical and human factors.

\textbf{Goals and Contributions.}
In this paper, we focus on hardware reverse engineering. Our goal is to discuss approaches to measure the complexity of reverse engineering with respect to both technical and human efforts. To this end, we first systematically survey reverse engineering methods and techniques described in the open literature. Based on our survey, we formulate several open research questions for quantification of reverse engineering. We then survey problem solving research and research on the acquisition of expertise, and briefly summarize what these approaches can provide to quantify the so-far neglected human factors in reverse engineering. Finally, we discuss how interdisciplinary research may be able to quantify the complexity of reverse engineering. In summary, our main contributions are:

\begin{itemize}
    \item {\bf Hardware Reverse Engineering Overview.}
    We systematically study hardware reverse engineering methods and techniques and provide a concise overview of the state of the art (\Section{hwre:section:hwre}).

   	\item {\bf Open Research Questions.}
    Based on the overview, we formulate several open research questions for reverse engineering with a focus on its quantification and  both technical and human factors (\Section{hwre:section:open_challenges}).

    \item {\bf Human Factors Quantification Overview.}
    To the best of our knowledge, we are the first to propose problem solving research and research on the acquisition expertise to quantify human factors in hardware reverse engineering (\Section{hwre:section:human_factor}). Finally, we discuss how interdisciplinary research with technical and humanistic perspectives may facilitate a sound quantification (\Section{hwre:section:quantification}).

\end{itemize}

%% file: section/hardware_reverse_engineering.tex

\section{Hardware Reverse Engineering}
\label{hwre:section:hwre}

In the following, we systematically survey hardware reverse engineering. To this end, we first detail the system model (\Section{hwre:section:system_model}). We then present diverse state of the art methods and techniques to analyze \acp{ASIC} (\Section{hwre:section:re:asic}) and \acp{FPGA} (\Section{hwre:section:re:fpga}) in order to retrieve the crucial gate-level netlists of a hardware design. Subsequently, we survey the state of the art in gate-level netlist reverse engineering (\Section{hwre:section:re:netlist}) which focusses on retrieval of high-level \ac{RTL} information (e.g. control unit or datapath components).
Finally, we formulate several open research questions with a particular focus on quantification of the complexity of reverse engineering (\Section{hwre:section:open_challenges}).

Note that a survey for anti reverse engineering techniques is out of the scope of this work, but the interested reader is referred to~\cite{book:hardware_obfuscation}.

\subsection{System Model}
\label{hwre:section:system_model}

We assume a reverse engineer with access to the flattened (placed and routed) gate-level netlist without any a priori knowledge of the design's internal workings. More precisely, the reverse engineer has no information of module hierarchies, synthesis options, or names of gates and signals.
The goal of the reverse engineer is to understand (parts of) the design's inner workings in order to perform another high-level application, i.e. to detect counterfeit products or to inject hardware Trojans.

The gate-level netlist can be obtained through several means in multiple real-world scenarios, i.e.
(1) chip-level reverse engineering (see \Section{hwre:section:re:asic}), or
(2) bitstream reverse engineering in case of \acp{FPGA} (see \Section{hwre:section:re:fpga}) or,
(3) directly from the layout in case of an untrusted (off-shore) foundry or from an \ac{IP} provider.

Note that this model is consistent with prior research on hardware security~\cite{alkabani:2007:usenix,chakraborty:2009:tcad,rostami:2014:ieee,bhunia:2014:ieee}.

\subsection{Chip-level Reverse Engineering}
\label{hwre:section:re:asic}

To access the gate-level netlist of an \ac{ASIC} post-manufacturing, chip-level reverse engineering has to be performed. Here, the goal is to deprocess the \ac{IC} which is embedded in the protective package. To this end, various steps are involved: (1) depackaging and mechanical preprocessing, (2) delayering and imaging, and (3) software post-processing~\cite{quadir:2016:jetc}.

\par{\bf Depackaging and Mechanical Preprocessing.}
The first goal of the chip-level reverse engineering step is to depackage the chip by use of wet-chemical or mechanical means. This is also called decapsulation. In particular, the die has to be protected from any harm, therefore, typically wet-chemical depackaging is chosen since the die is protected by a seal-layer from the front side, i.e. often an $SiO_2$ passivation. Note that the backside usually offers enough silicon in the bulk to withstand careful depackaging processes. Additionally, bonding wires are of special interest during any semi-invasive attacks since they connect the embedded die to the package pins.

\par{\bf Delayering and Imaging.}
Once the die is fully recovered, the \ac{IC} is delayered and digitized by optical means, i.e. a \ac{SEM} or \ac{FIB}. The delayering process can involve a combination of different wet-chemical, plasma-etching, and mechanical polishing steps. Note that especially during these steps the handling of the equipment results in improved quality of the results. In particular, planarization of the current layer with a large \textit{surface-to-thickness} ratio is challenging in practice. Also knowing the \ac{ROI} is beneficial as the planar surface can be reduced significantly. In such cases, the reverse engineer can pinpoint his \ac{ROI} while neglecting the rest of the chip~\cite{ches:2015:aeshaystack}.

Furthermore, every chip has different chip manufacturing processes due to cost optimizations or technology node requirements. Therefore different conductors, semiconductors and dielectrics have to be investigated and selectively removed without destroying functional information of the \ac{IC}~\cite{quadir:2016:jetc}. For modern, nanoscale technologies, it is essential to have the necessary equipment to approximate or measure remaining layer thickness and assess delayering quality.

\begin{figure}[!htb]
\centering
  \includegraphics[width=0.25\textwidth]{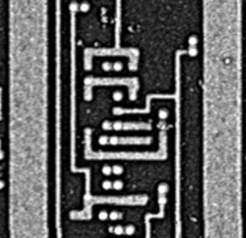}
  \caption{Example of metal 1 layer is shown. Brightness allows to distinguish between wires, vias, and the \ac{SOD}. The brighter dots are vias between Metal 1 and Metal 2.}
  \label{fig:normalreverse}
  \vspace{-5pt}
\end{figure}

In state of the art reverse engineering, digitalizing and imaging is performed using a \ac{SEM} or \ac{FIB}. Since modern technology sizes hit the diffraction limit of optical microscopes, more advanced visualizing tools are mandatory. On the one hand such modern equipment is costly, on the other hand it results in smaller images.  During image acquiring, a brightness yield from the metals to the vias and a brightness difference to the background is created due to different substance (electrical-) properties. A clear brightness yield from the \ac{SEM}/\ac{FIB} images is beneficial for the post-processing as it allows to distinguish between vias, wires and \ac{SOD}, see Figure~\ref{fig:normalreverse}.

In summary, a good understanding of the physics of the processes  and the necessary equipment is mandatory to achieve an adequate delayering quality, but also personal safety.
Handling of highly concentrated acids (e.g., hydrofluoric acid HF), should only be done with the necessary knowledge in chemistry. Sometimes a single scratch from dust in the laboratory means the end of an \ac{IC} sample.

\newpage
\par{\bf Software Post-processing.}
In order to generate a functional chip representation from the digitized tile images of the previous step, the tile images have to be stitched and vectorized. As every \ac{IC} is built from a standard-cell library, every cell in this library has to be recognised manually and its functional interpretation extracted once. Once the cell is identified, it can be automatically detected in the whole \ac{ROI} by means of image processing. Finally, with the standard cell instances in the back-end-of-line and the post-processed metal lines in vector representation the functional interpretation of the a \ac{ROI} can be extracted as a netlist. Note this is a tedious and repetitive task that can be (semi-) automated to support the reverse engineer.

\subsection{\ac{FPGA} Bitstream Reverse Engineering}
\label{hwre:section:re:fpga}

In order to access gate-level netlists of \acp{FPGA}, a reverse engineer has to analyze the configuration bitstream file that defines its behavior. To this end, the reverse engineer has to (1) access the bitstream, (2) decrypt the bitstream (in case bitstream encryption is deployed), and (3) perform reverse engineering of the proprietary bitstream file format to retrieve the netlist~\cite{quadir:2016:jetc}. Note that the following description focuses on the market-dominating \ac{SRAM}-based \acp{FPGA} technology, for other technologies the interested reader is referred to Wanderley~\etal~\cite{book:security_trends_fpga:chapter2}.

\par{\bf Bitstream Access.}
Due to the underlying \ac{SRAM} technology, \ac{SRAM}-based \acp{FPGA} require external non-volatile memory such as flash to store the bitstream. Hence, a reverse engineer can either access the non-volatile memory and dump its content, or wire-tap the communication between \ac{FPGA} and non-volatile memory upon boot-up, cf.~\cite{jcen:2016:swierczynski}

\par{\bf Bitstream Decryption.}
In order to provide confidentiality of the bitstream, \ac{FPGA} manufacturers deployed a bitstream encryption scheme for various device series using strong cryptographic primitives. However, several works have demonstrated that various series are vulnerable to side-channel attacks which recover the secret encryption keys~\cite{moradi:2011:ccs,moradi:2013:fpga}. Thus even if bitstream encryption is deployed, the bitstream can be decrypted for the majority of series. Note that most low-cost series do not offer bitstream encryption at all.

\par{\bf Bitstream Reverse Engineering.}
Since the bitstream file format is proprietary, a reverse engineer has to analyze the file format in order to transform the (decrypted) bitstream into its readable gate-level netlist description. To this end, several works developed automated file format reverse engineering strategies to recover (partial) netlist information, cf.~\cite{book:security_trends_fpga:chapter2}.

\subsection{Gate-level Netlist Reverse Engineering}
\label{hwre:section:re:netlist}

After we specified how a reverse engineer can access the gate-level netlist for \acp{ASIC} and \acp{FPGA} designs, we now provide an overview of publicly documented reverse engineering techniques to retrieve high-level information (e.g., control units or hierarchy information of submodules).

Chisholm~\etal~\cite{dt:1999:chisholm} presented a workflow on how to reverse engineer module-level descriptions from gate-level netlists, addressing the synergy of the human analyst's creativity and the computer's ability to solve repetitive tasks. In a case study, Hansen~\etal~\cite{hansen:1999:dt} described several best-practices for a human analysts to reverse engineer gate-level netlists.
Shi~\etal~\cite{shi:2010:iscas} evaluated a technique to automatically reverse engineer circuitry that control units, i.e. \ac{FSM} from gate-level netlists. Meade~\etal~\cite{meade:2016:aspdac} extended this technique in order to retrieve the state transition function for the reverse engineered \acp{FSM}. In further work, Meade~\etal~\cite{meade:2016:iscas} developed a technique to separate control unit registers from datapath registers.
In order to automatically reverse engineer functional submodules in a larger hardware design, diverse techniques have been developed based on Boolean function analysis~\cite{shi:2012:iscit}, pattern mining of simulation traces and model checking~\cite{li:2012:host}, module boundary identification~\cite{subramanyan:2013:date, subramanyan:2013:tect}, and word-level structure identification~\cite{li:2013:host}.
Since functional identification of subcircuits requires to find the correct matching between known subcircuits and the subcircuit under inspection, a reverse engineer has to find the correct input permutation. To avoid this computationally expensive task, Gasc\'{o}n~\etal
\cite{gascon:2014:fmcad} addressed this problem with a template-based
solution.

\subsection{Open Challenges for Quantification}
\label{hwre:section:open_challenges}

In the previous sections, we highlighted several scenarios in which a reverse engineer can access gate-level netlists and we provided a concise background on how these netlists are reverse engineered. Even though several automated techniques and best-practices for a human analyst have been described in the literature so far, there are still open challenges with respect to quantification of the reverse engineering process: (1) automation of reverse engineering techniques, and (2) quantification of the remaining non-automated sensemaking by human analysts.

\par{\bf Automated Reverse Engineering Techniques.}
Future research in the field of hardware reverse engineering should focus on further automated techniques to retrieve high-level information from gate-level netlists.
On the one hand novel automated techniques investigate which information can be algorithmically extracted, and on the other hand they simultaneously provide a fine-grained quantification of the time complexity.

\par{\bf Quantification of Human Factors.}
Since reverse engineering always involves human analysts, metrics for reverse engineering and obfuscation which solely focus on technical aspects are apparently not adequate. Thus a key challenge for future research is to quantify the human factor in reverse engineering.

%% file: section/challenges_human_factor.tex

\section{Problem Solving and Expertise Research}
\label{hwre:section:human_factor}

We now present problem solving and expertise research in the context of hardware reverse engineering, in particular for quantification of gate-level netlist reverse engineering.
First, we highlight why this reverse engineering task is a problem solving process (\Section{hwre:section:setting}). We then provide a general introduction to problem solving research and research on the acquisition of expertise (\Section{hwre:section:learning_problem_solving}). Finally, we propose how the human factor can be quantified using the aforementioned psychological research fields (\Section{hwre:section:quantification}).

\subsection{Setting-A Learning Perspective}
\label{hwre:section:setting}

As stated in~\Section{hwre:section:system_model}, we assume a reverse engineer with access to a gate-level netlist and the goal to understand parts of the design's inner workings. To this end, the analyst chooses actions which reduce the difference between the \textit{initial} state (no high-level information) and the \textit{goal} state (design's inner workings successfully extracted). During this process the person draws on prior knowledge (e.g., knowledge from past instances of gate-level netlist reverse engineering or textbooks). Thus, knowledge generated during reverse engineering can be utilized in future attempts.

\par{\bf Perspectives on the Human Factor.}
This setting points out two separate but intertwined mechanisms: (1) gate-level netlist reverse engineering can be viewed as a problem solving process, and (2) reverse engineers can acquire new knowledge or skills and store them in long-term memory ~\cite{Atkinson.1968}. In particular, reverse engineers gain expertise by performing reverse engineering repeatedly in different contexts, i.e. formal (e.g., school, university) and/or informal (e.g., learning or training on-the-job, self-study, exchange with peers) educational settings~\cite{Werquin.2007}.

Both mechanisms, problem solving and acquisition of expertise, also describe the arms race of reverse engineering and obfuscation, since reverse engineers are able to break obfuscation strategies and use their gained experience for future reverse engineering attempts. Designers then have to implement a new obfuscation, which presents a novel problem to reverse engineers.

\subsection{Problem Solving and Expertise Research}
\label{hwre:section:learning_problem_solving}

Based on the learning perspective in the previous section, we survey problem solving and expertise research for a general audience. While problem solving research focuses on problem properties and adaptation of strategies to overcome obstacles, expertise research conceptualizes the development of knowledge required for successful reverse engineering, changes in problem solving strategies with accumulating experience, mental representations of problems and increasing automation of complex and initially effortful behaviors (experts vs. novices) ~\cite{Ollinger.2017}, ~\cite{Nokes.2010}.

\par{\bf Problem Solving Research.} Problem solving can be defined as a sequence of directed cognitive operations that are employed in a situation (the problem) where the individual does not possess a suitable routine operation that allows a transition from a given initial state to the desired goal state. This situation is termed problem ~\cite{Mayer.2002, Ollinger.2017}. During problem solving, knowledge is manipulated in order to attain the desired goal state.
Due to different prior knowledge and problem solving skills, a situation might pose a problem to one person, but not to another. Further, as soon as a person has solved a problem and is able to fully reproduce the solution schema, the situation loses its problem character and simply represents a task to this individual~\cite{Hussy.1998}. Thus, learning from problem solving as an ongoing process should be taken into consideration as well.

\par{\bf Expertise Research.} Ongoing experience within one field, combined with deliberate practice~\cite{Ericsson.1993} results in acquisition of expertise. Deliberate practice refers to a specific practice or training activity in which a person willingly and repeatedly produces an action (often under supervision). The trainee receives feedback on the quality of the production of the action with the ultimate goal to improve performance~\cite{Ericsson.1993, Ericsson.2010}. After an extensive amount of practice a person is capable of repeatedly exhibiting superior performance with minimal variation. Note that this separates the acquisition of expertise from the acquisition of (everyday) skills, which eventually reaches an autonomous stage where performance will no longer improve~\cite{Ericsson.2003}.
In contrast to novices, experts perform superior due to their improved working memory, cf.~\cite{Atkinson.1968}, which allows them to process great amount of information at a time. For example, chess masters are able to quickly perceive and evaluate complex configurations and choose promising options for further moves~\cite{Groot.1978}. Due to repeated problem solving, experts also possess \textit{problem-schemas}, which allow them to identify the deep structure of a problem, retrieve multiple solutions and select the best solution~\cite{Jonassen.2000, Chi.1981}.

\section{Open Challenges: Quantification of Human Factors}
\label{hwre:section:quantification}

As discussed in \Section{hwre:section:open_challenges}, quantification of human factors is an open challenge in hardware security research. For the quantification using problem solving and expertise research described in~\Section{hwre:section:human_factor}, several aspects can be investigated such as the reverse engineering process itself or how the human experiment is arranged.

In the following, we provide novel interdisciplinary perspectives that systematically capture the different aspects of human factor quantification for reverse engineering. First, we propose two dichotomies which can guide quantification of reverse engineering , namely (1) \textit{process vs. result}, and (2) \textit{human vs. task} (\Section{hwre:section:dichotomy}). Second, we discuss possible research designs and methods of data collection to investigate the human factor (\Section{hwre:section:research_design}).

\subsection{Dichotomies for Human Factor Quantification}
\label{hwre:section:dichotomy}

We present a systematic overview of quantifiable aspects arranged in two dichotomies. Note that both dichotomies are not mutually exclusive, but rather represent different perspectives of gate-level netlist reverse engineering.

\subsubsection{Dichotomy: Process vs. Results Quantification}
For the quantification of human factors using problem solving and expertise research as described in~\Section{hwre:section:human_factor}, it appears reasonable to distinguish between quantification of the process itself and quantification of its results.

\par{\bf Process Quantification Dimension.}
The primary scope of the \textit{process quantification dimension} is not \textit{if} analysts are able to reverse engineer a netlist, but \textit{how} they solve problems they encounter. Usually, analysts identify high-level steps and define a set of main goals to complete. These steps represent meaningful units that guide analysts and pose specific challenges to them. In order to complete these steps, analysts might need to employ a number of different strategies.

The process dimension also focuses on learning gains and the time required to complete a given task, i.e. how fast can an analyst learn to master a task and how long does it take to accomplish the task of a given complexity?
These processes change over time as individuals repeatedly encounter problems of a similar topography and their actions become automated~\cite{Ericsson.2006}).

With regard to expertise and its acquisition, characterization of expertise-specific problem solving strategies and problem representations is expedient for quantification, since experts have different ways of perceiving problems and employ qualitatively different problem solving strategies due to their superior knowledge organization compared to novices (this has been shown in domains like chess~\cite{Chase.1973}, physics~\cite{Chi.1981}, symbolic drawings in electrical engineering~\cite{Egan.1979} or computer programming~\cite{McKeithen.1981}).

\par{\bf Result Quantification Dimension.} The primary scope of the \textit{result quantification dimension} is to investigate whether analysts were \textit{successful} in reverse engineering and \textit{what they have learned} during problem solving, i.e. what new knowledge or skills they acquired.
Analysts acquire new (domain-specific) problem solving strategies or reach an improved proficiency in utilizing already learned strategies.

Considering the acquisition of expertise, it is also important to assess whether analysts can reproduce their solution on similar problems. This asserts whether or not a problem class of challenges still poses a problem to the analysts. Moreover it is necessary to investigate to what extend analysts can transfer their knowledge about the solution of a problem solved to a structurally similar problem.

\subsubsection{Dichotomy: Quantification of Human vs. Task Properties}
\label{hwre:section:DychHumTask}
Another dichotomy for the quantification is to distinguish between properties of the analyst and properties of the task, i.e. the hardware design.

\par{\bf Human Property Dimension.}
The primary scope of the \textit{human property dimension} is the analysis of characteristics required for reverse engineering, e.g. domain knowledge, technical skills, and broader human traits, such as general intelligence. These factors determine how and to what result an analyst is able to solve a reverse engineering task. The comparison of such capacities between subjects may yield a more sophisticated understanding of characteristics to distinguish experts from novices, and the identification of useful predictors (and less relevant factors, or even obstacles) for successful reverse engineering.

\par{\bf Task Property Dimension.}
The primary scope of the \textit{task property dimension} is to analyze the characteristics of the target hardware design, i.e. the amount of gates and the complexity of their interconnections.

Whereas the difficulty of so-called simple problems (e.g. The Tower of Hanoi ~\cite{Anderson.1993}) merely determines the amount of time required to solve it (due to an increasing number of incremental steps or iterations required), increasing the difficulty of complex problems ~\cite{Funke.2012} (i.e. the amount of relevant information to be considered or processed simultaneously) may further diminish the problem solving performance, i.e. the quality of the solution. Analyzing which components of a netlist should be considered simple, and which complex, is key to quantifying the human factor in reverse engineering.

Further, since analysts might use tools to transform the original gate-level netlists into a graph-based representation to conduct visual pattern matching search strategies, research on \textit{insight problems}~\cite{Pretz.2003} might indicate design characteristics that facilitate or hinder reverse engineering when using visualizations.

\subsection{Research Designs, Data Collection, and Challenges}
\label{hwre:section:research_design}

In addition to the quantifiable aspects in the previous section, we now present (1) aspects for research designs, (2) methods of data collection used to investigate the human factor, and (3) challenges for future research.
In order to quantify the human factor, researchers collect data on the process and outcome of reverse engineering attempts and which human characteristics and task influence reverse engineering success. By choosing between different research designs and methods of data collection,the researcher selects which parts of reality are under investigation and which are excluded. Therefore, carefully choosing designs and methods of data collection with regard to the research question and their respective advantages and disadvantages is an important task.

\subsubsection{Research Designs: Laboratory vs. Field}

Research on the human factor can be either carried out in laboratory studies or in the field. Field studies are observations at the places where analysts \textit{naturally} perform hardware reverse engineering. This allows gaining insight into the complexity of the processes in their respective context. Conversely, in laboratory experiments researchers control the context and observe single aspects in great detail and with reduced external influences as compared to observations in field. These studies take place at the researchers' laboratories. Please note that the term \textit{laboratory} refers to the artificiality of the context and must not be confused with the reverse engineer's laboratory - which constitutes the site of a field study. Here, reverse engineering is carried out under \textit{artificial} conditions. However, researchers may underestimate or mischaracterize such processes as they impose unrealistic boundary conditions, restrict access to resources analysts might normally use, or fail to capture relevant strategies not available in a laboratory setting. The strict control of context, however, is key to investigate and isolate the effect or role of particular variables. Research in the laboratory requires researchers to use a formal description of the reverse engineering process (see \Section{hwre:section:Challenges}.

\subsubsection{Data Collection}
\

\par{\bf Behavioral Data.}
Behavior observations such as log-files from human-computer-interaction, eye-tracking, screen captures, or videographs allow a detailed analysis of actions and strategies and their respective development over time. Behavioral data is not affected by shortcomings concerning memory, introspection, or response biases. Meaningfully reconstructing behavioral sequences from logfiles, however, requires a sophisticated system to be set up a priori. Behaviors not expected to occur by the researcher may simply not be reconstructible from such automated recordings, rendering them arguably incomplete or even useless.

\par{\bf Verbal Data.}
Having analysts verbalize their thoughts while they are problem solving (e.g., \textit{think-aloud}) allows insights into mental models, deliberations and intentions behind the strategies employed (e.g., a sequence of goals)~\cite{COOKE1994801}. An alternative to think-aloud is \textit {stimulated recall}~\cite{COOKE1994801}. With this data collection technique, the problem solving process itself is not verbalized during its course, but the process is recorded. After problem solving, the problem solvers are presented a section of their problem solving behavior and are asked to explain their actions. While information revealed that way is valuable to understand a phenomenon, the quality of such self-report data may be limited when respondents are unwilling or unable to provide an accurate account \cite{Ericsson.1980, Nisbett.1977}.

\subsubsection{Challenges}
\label{hwre:section:Challenges}

\par{\bf Process Description.}
A major challenge for research is the lack of a formal description of how analysts carry out reverse engineering. Understanding the structure of a problem is an important prerequisite in order to investigate the cognitive processes involved in solving the problem \cite{Ollinger.2017}. Applying a formalized description during research on reverse engineering makes results of different research groups comparable, facilitates an integration of findings and allows meaningful research synthesis.

\par{\bf Sampling.}
Meaningful research on reverse engineering requires sampling of subjects trained in reverse engineering which is a highly domain-specific process which presumably only relatively few people are capable of. This dramatically reduces the population to draw samples from. Among those, a substantial proportion will be unwilling to follow an invitation to a university laboratory (as their reverse engineering usually pursues illegitimate purposes), and some might be bound by contracts or other agreements that prevents them from participating. In addition, building contact to the remaining potential participants and thus recruiting research participants may be challenging, and researchers interested in studying the human factor in reverse engineering are advised to pool their resources.

%% file: section/conclusion.tex

\section{Conclusion}
\label{hwre:section:conclusion}

Both industry and academia have been dealing with hardware reverse engineering for several decades. Although reverse engineering serves various legitimate and illegitimate applications, quantification of its complexity is an unsolved problem so far. However, this quantification is crucial in order to provide reasonable threat estimation and to develop sound countermeasures to mitigate risks posed by reverse engineering.

In this work, we first systematically analyzed the state of the art in hardware reverse engineering and identified two major open research directions: (1) automation of technical factors, and (2) quantification of the remaining non-automated sensemaking conducted by human analysts. We then surveyed problem solving research and research on the acquisition of expertise for a general audience which facilitates quantification of decisive human factors.
Finally, by broadening the scope of reverse engineering through combination of technical and human-centered perspectives, we provide suggestions for future research directions to holistically capture the complexity of hardware reverse engineering.

We believe that our insights on hardware reverse engineering and its open challenges will help other researchers in finding new ways to move the state of the art in this area forward.

%% file: main.bbl
\begin{thebibliography}{10}
\providecommand{\url}[1]{#1}
\csname url@samestyle\endcsname
\providecommand{\newblock}{\relax}
\providecommand{\bibinfo}[2]{#2}
\providecommand{\BIBentrySTDinterwordspacing}{\spaceskip=0pt\relax}
\providecommand{\BIBentryALTinterwordstretchfactor}{4}
\providecommand{\BIBentryALTinterwordspacing}{\spaceskip=\fontdimen2\font plus
\BIBentryALTinterwordstretchfactor\fontdimen3\font minus
  \fontdimen4\font\relax}
\providecommand{\BIBforeignlanguage}[2]{{%
\expandafter\ifx\csname l@#1\endcsname\relax
\typeout{** WARNING: IEEEtran.bst: No hyphenation pattern has been}%
\typeout{** loaded for the language `#1'. Using the pattern for}%
\typeout{** the default language instead.}%
\else
\language=\csname l@#1\endcsname
\fi
#2}}
\providecommand{\BIBdecl}{\relax}
\BIBdecl

\bibitem{smc:1985:rekoff}
M.~G. Rekoff, ``{On Reverse Engineering},'' \emph{{IEEE Transactions on
  Systems, Man, and Cybernetics}}, no.~2, pp. 244--252, 1985.

\bibitem{it:2012:willems}
C.~Willems and F.~C. Freiling, ``Reverse code engineering - state of the art
  and countermeasures,'' \emph{it - Information Technology}, vol.~54, no.~2,
  pp. 53--63, 2012.

\bibitem{isc:2003:oorschot}
P.~C.~V. Oorschot, ``{Revisiting Software Protection},'' in \emph{ISC 2003.
  LNCS}.\hskip 1em plus 0.5em minus 0.4em\relax Springer, 2003, pp. 1--13.

\bibitem{book:hardware_obfuscation:chapter1}
{B. Shakya \etal}, ``{Introduction to Hardware Obfuscation: Motivation, Methods
  and Evaluation},'' in \emph{Hardware Protection through Obfuscation}.\hskip
  1em plus 0.5em minus 0.4em\relax Springer, 2017, ch.~1, pp. 3--32.

\bibitem{quadir:2016:jetc}
{S. E. Quadir \etal}, ``A survey on chip to system reverse engineering,''
  \emph{{JETC}}, vol.~13, no.~1, pp. 6:1--6:34, 2016.

\bibitem{subramanyan:2013:tect}
{P. Subramanyan et al.}, ``{Reverse Engineering Digital Circuits Using
  Structural and Functional Analyses},'' \emph{{IEEE} Trans. Emerging Topics
  Comput.}, vol.~2, no.~1, pp. 63--80, 2014.

\bibitem{ieee:2014:guin}
{U. Guin \etal}, ``{Counterfeit Integrated Circuits: A Rising Threat in the
  Global Semiconductor Supply Chain},'' \emph{Proceedings of the IEEE}, vol.
  102, no.~8, pp. 1207--1228, 2014.

\bibitem{bhunia:2014:ieee}
{S. Bhunia \etal}, ``{Hardware Trojan Attacks: Threat Analysis and
  Countermeasures},'' \emph{Proceedings of the IEEE}, vol. 102, no.~8, pp.
  1229--1247, 2014.

\bibitem{vijayakumar:2017:tifs}
{A. Vijayakumar \etal}, ``{Physical Design Obfuscation of Hardware: {A}
  Comprehensive Investigation of Device and Logic-Level Techniques},''
  \emph{{IEEE} Trans. Information Forensics and Security}, vol.~12, no.~1, pp.
  64--77, 2017.

\bibitem{spectrum:2000:kumagai}
J.~Kumagai, ``{Chip detectives},'' \emph{{IEEE Spectrum}}, vol.~37, no.~11, pp.
  43--48, 2000.

\bibitem{torrance:2009:ches}
R.~Torrance and D.~James, ``{The State-of-the-Art in IC Reverse Engineering},''
  in \emph{CHES}.\hskip 1em plus 0.5em minus 0.4em\relax Springer, 2009, pp.
  363--381.

\bibitem{book:hardware_obfuscation}
{D Forte \etal}, \emph{{Hardware Protection through Obfuscation}},
  1st~ed.\hskip 1em plus 0.5em minus 0.4em\relax {Springer}, 2017.

\bibitem{alkabani:2007:usenix}
Y.~Alkabani and F.~Koushanfar, ``{Active Hardware Metering for Intellectual
  Property Protection and Security},'' in \emph{{USENIX Security Symposium}},
  2007.

\bibitem{chakraborty:2009:tcad}
R.~S. Chakraborty and S.~Bhunia, ``{HARPOON: An Obfuscation-Based SoC Design
  Methodology for Hardware Protection},'' \emph{{IEEE} Trans. on {CAD} of
  Integrated Circuits and Systems}, vol.~28, no.~10, pp. 1493--1502, 2009.

\bibitem{rostami:2014:ieee}
M.~Rostami, F.~Koushanfar, and R.~Karri, ``{A Primer on Hardware Security:
  Models, Methods, and Metrics},'' \emph{Proceedings of the {IEEE}}, vol. 102,
  no.~8, pp. 1283--1295, 2014.

\bibitem{ches:2015:aeshaystack}
C.~Kison, J.~Frinken, and C.~Paar, ``Finding the aes bits in the haystack:
  Reverse engineering and sca using voltage contrast,'' in \emph{CHES}.\hskip
  1em plus 0.5em minus 0.4em\relax Springer, 2015, pp. 641--660.

\bibitem{book:security_trends_fpga:chapter2}
{E. Wanderley \etal}, \emph{{Security FPGA Analysis}}.\hskip 1em plus 0.5em
  minus 0.4em\relax Springer, 2011, pp. 7--46.

\bibitem{jcen:2016:swierczynski}
{P. Swierczynski \etal}, ``{Interdiction in Practice---Hardware Trojan Against
  a High-Security USB Flash Drive},'' \emph{{Journal of Cryptographic
  Engineering}}, pp. 1--13, 2016.

\bibitem{moradi:2011:ccs}
{A. Moradi \etal}, ``{On the Vulnerability of FPGA Bitstream Encryption against
  Power Analysis Attacks: Extracting Keys from Xilinx Virtex-II FPGAs},'' in
  \emph{CCS}, 2011, pp. 111--124.

\bibitem{moradi:2013:fpga}
------, ``{Side-channel Attacks on the Bitstream Encryption Mechanism of Altera
  Stratix II: Facilitating Black-box Analysis Using Software
  Reverse-engineering},'' in \emph{FPGA}, 2013, pp. 91--100.

\bibitem{dt:1999:chisholm}
{G. H. Chisholm \etal}, ``{Understanding Integrated Circuits},'' \emph{{IEEE}
  Design {\&} Test of Computers}, vol.~16, no.~2, pp. 26--37, 1999.

\bibitem{hansen:1999:dt}
{M. C. Hansen \etal}, ``{Unveiling the {ISCAS-85} Benchmarks: {A} Case Study in
  Reverse Engineering},'' \emph{{IEEE} Design {\&} Test of Computers}, vol.~16,
  no.~3, pp. 72--80, 1999.

\bibitem{shi:2010:iscas}
{Y. Shi \etal}, ``A highly efficient method for extracting fsms from flattened
  gate-level netlist,'' in \emph{{ISCAS}}, 2010, pp. 2610--2613.

\bibitem{meade:2016:aspdac}
{T. Meade \etal}, ``{Netlist Reverse Engineering for High-Level Functionality
  Reconstruction},'' in \emph{ASP-DAC}, 2016, pp. 655--660.

\bibitem{meade:2016:iscas}
------, ``{Gate-level Netlist Reverse Engineering for Hardware Security:
  Control Logic Register Identification},'' in \emph{ISCAS}, 2016, pp.
  1334--1337.

\bibitem{shi:2012:iscit}
{Y. Shi \etal}, ``Extracting functional modules from flattened gate-level
  netlist,'' in \emph{{ISCIT}}, 2012, pp. 538--543.

\bibitem{li:2012:host}
{W. Li \etal}, ``Reverse engineering circuits using behavioral pattern
  mining,'' in \emph{{HOST}}, 2012, pp. 83--88.

\bibitem{subramanyan:2013:date}
{P. Subramanyan \etal}, ``{Reverse Engineering Digital Circuits Using
  Functional Analysis},'' in \emph{{DATE}}, 2013, pp. 1277--1280.

\bibitem{li:2013:host}
{W. Li \etal}, ``Wordrev: Finding word-level structures in a sea of bit-level
  gates,'' in \emph{{HOST}}, 2013, pp. 67--74.

\bibitem{gascon:2014:fmcad}
{A. Gasc{\'{o}}n et al.}, ``Template-based circuit understanding,'' in
  \emph{FMCAD}, 2014, pp. 83--90.

\bibitem{Atkinson.1968}
R.~C. Atkinson and R.~M. Shiffrin, ``Human memory: A proposed system and its
  control processes1,'' in \emph{The psychology of learning and motivation},
  K.~W. Spence, J.~{Taylor Spence}, and J.~T. Spence, Eds.\hskip 1em plus 0.5em
  minus 0.4em\relax New York: {Academic Press}, 1968, vol.~2, pp. 89--195.

\bibitem{Werquin.2007}
P.~Werquin, ``Terms, concepts and models for analysing the value of recognition
  programmes: Rnfil - third meeting of national representatives and
  international organisations,'' Vienna, Austria.

\bibitem{Ollinger.2017}
M.~{\"O}llinger, ``Probleml{\"o}sen,'' in \emph{Allgemeine Psychologie},
  J.~M{\"u}sseler and M.~Rieger, Eds.\hskip 1em plus 0.5em minus 0.4em\relax
  Berlin, Heidelberg: {Springer Berlin Heidelberg}, 2017, pp. 587--618.

\bibitem{Nokes.2010}
T.~J. Nokes, C.~D. Schunn, and M.~Chi, ``Problem solving and human expertise,''
  in \emph{International Encyclopedia of Education (Third Edition)},
  P.~Peterson, E.~Baker, and B.~McGaw, Eds.\hskip 1em plus 0.5em minus
  0.4em\relax Oxford: Elsevier, 2010, pp. 265--272.

\bibitem{Mayer.2002}
R.~E. Mayer, ``A taxonomy for computer-based assessment of problem solving,''
  \emph{Computers in Human Behavior}, vol.~18, no.~6, pp. 623--632, 2002.

\bibitem{Hussy.1998}
W.~Hussy and H.~Selg, \emph{Denken und Probleml{\"o}sen}, 2nd~ed., ser.
  Urban-Taschenb{\"u}cher.\hskip 1em plus 0.5em minus 0.4em\relax Stuttgart:
  Kohlhammer, 1998, vol. 557.

\bibitem{Ericsson.1993}
K.~A. Ericsson, R.~T. Krampe, and C.~Tesch-R{\"o}mer, ``The role of deliberate
  practice in the acquisition of expert performance,'' \emph{Psychological
  Review}, vol. 100, no.~3, pp. 363--406, 1993.

\bibitem{Ericsson.2010}
K.~A. Ericsson and T.~J. Towne, ``Expertise,'' \emph{Wiley interdisciplinary
  reviews. Cognitive science}, vol.~1, no.~3, pp. 404--416, 2010.

\bibitem{Ericsson.2003}
K.~A. Ericsson, ``The acquisition of expert performance as problem solving:
  Construction and modification of mediating mechanisms through deliberate
  practice,'' in \emph{The psychology of problem solving}, J.~E. Davidson and
  R.~J. Sternberg, Eds.\hskip 1em plus 0.5em minus 0.4em\relax Cambridge, UK
  and New York: {Cambridge University Press}, 2003, pp. 31--83.

\bibitem{Groot.1978}
A.~D.~d. Groot, \emph{Thought and Choice in Chess}, 2nd~ed., ser. Psychological
  Studies.\hskip 1em plus 0.5em minus 0.4em\relax Berlin/Boston: {De Gruyter}
  and {De Gruyter Mouton}, 1978, vol.~4.

\bibitem{Jonassen.2000}
D.~H. Jonassen, ``Toward a design theory of problem solving,''
  \emph{Educational Technology Research and Development}, vol.~48, no.~4, pp.
  63--85, 2000.

\bibitem{Chi.1981}
M.~T.~H. Chi, P.~J. Feltovich, and R.~Glaser, ``Categorization and
  representation of physics problems by experts and novices,'' \emph{Cognitive
  Science}, vol.~5, no.~2, pp. 121--152, 1981.

\bibitem{Ericsson.2006}
K.~A. Ericsson, ``The influence of experience and deliberate practice on the
  development of superior expert performance,'' in \emph{The Cambridge Handbook
  of Expertise and Expert Performance}, K.~A. Ericsson, Ed.\hskip 1em plus
  0.5em minus 0.4em\relax {Cambridge University Press}, 2006, pp. 683--704.

\bibitem{Chase.1973}
S.~G. Chase and H.~Simon, ``Perception in chess,'' \emph{Cognitive Psychology},
  no.~4, pp. 55--81, 1973.

\bibitem{Egan.1979}
D.~E. Egan and B.~J. Schwartz, ``Chunking in recall of symbolic drawings,''
  \emph{Memory {\&} Cognition}, vol.~7, no.~2, pp. 149--158, 1979.

\bibitem{McKeithen.1981}
K.~B. McKeithen, J.~S. Reitman, H.~H. Rueter, and S.~C. Hirtle, ``Knowledge
  organization and skill differences in computer programmers,'' \emph{Cognitive
  Psychology}, vol.~13, no.~3, pp. 307--325, 1981.

\bibitem{Anderson.1993}
J.~R. Anderson, ``Problem solving and learning,'' \emph{American Psychologist},
  vol.~48, no.~1, pp. 35--44, 1993.

\bibitem{Funke.2012}
J.~Funke, ``Complex problem solving,'' in \emph{Encyclopedia of the Sciences of
  Learning}, N.~M. Seel, Ed.\hskip 1em plus 0.5em minus 0.4em\relax Boston, MA:
  {Springer US}, 2012, pp. 682--685.

\bibitem{Pretz.2003}
J.~E. Pretz, A.~J. Naples, and R.~J. Sternberg, ``Recognizing, defining, and
  representing problems,'' in \emph{The psychology of problem solving}, J.~E.
  Davidson and R.~J. Sternberg, Eds.\hskip 1em plus 0.5em minus 0.4em\relax
  Cambridge, UK and New York: {Cambridge University Press}, 2003, pp. 3--30.

\bibitem{COOKE1994801}
N.~J. Cooke, ``Varieties of knowledge elicitation techniques,''
  \emph{International Journal of Human-Computer Studies}, vol.~41, no.~6, pp.
  801 -- 849, 1994.

\bibitem{Ericsson.1980}
K.~A. Ericsson and H.~A. Simon, ``Verbal reports as data,'' \emph{Psychological
  Review}, vol.~87, no.~3, pp. 215--251, 1980.

\bibitem{Nisbett.1977}
R.~E. Nisbett and T.~D. Wilson, ``Telling more than we can know: Verbal reports
  on mental processes,'' \emph{Psychological Review}, vol.~84, no.~3, pp.
  231--259, 1977.

\end{thebibliography}
